\documentclass[10pt, conference, letterpaper]{IEEEtran}

\usepackage[nocompress]{cite}
\usepackage{booktabs}
\usepackage{amsmath,amssymb,amsthm}
\usepackage{todonotes}

\usepackage[caption=false,font=footnotesize]{subfig}

\usepackage{graphicx}
\usepackage[noend]{algpseudocode}
\usepackage{algorithm}
\usepackage{multirow}
\usepackage{array}
\usepackage{url}
\usepackage{lipsum}
\usepackage[utf8]{inputenc}		

\usepackage[draft,footnote,nomargin]{fixme}
\fxusetheme{color}

\graphicspath{{figures/}}

\begin{document}

\title{Specifying and Analyzing Virtual Network Services Using Queuing Petri Nets}

\author{\IEEEauthorblockN{Stefan Schneider, Arnab Sharma, Holger Karl, Heike Wehrheim}
\IEEEauthorblockA{Paderborn University, Paderborn, Germany\\
\{stefan.schneider, arnab.sharma, holger.karl, wehrheim\}@uni-paderborn.de}}

\maketitle

\begin{abstract}
For optimal placement and orchestration of network services, it is crucial that their structure and semantics are specified clearly and comprehensively and are available to an orchestrator.
Existing specification approaches are either ambiguous or miss important aspects regarding the behavior of virtual network functions~(VNFs) forming a service.
We propose to formally and unambiguously specify the behavior of these functions and services using Queuing Petri Nets~(QPNs).
QPNs are an established method that allows to express queuing, synchronization, stochastically distributed processing delays, and changing traffic volume and characteristics at each VNF.
With QPNs, multiple VNFs can be connected to complete network services in any structure, even specifying bidirectional network services containing loops.

We propose a tool-based workflow that supports the specification of network services and the automatic generation of corresponding simulation code to enable an in-depth analysis of their behavior and performance. In a case study, we show how developers can benefit from analysis insights, e.g., to anticipate the impact of different service configurations.
We also discuss how management and orchestration systems can benefit from our clear and comprehensive specification approach and its extensive analysis possibilities, leading to better placement of VNFs and improved Quality of Service.


\end{abstract}

\section{Introduction}\label{sec:introduction}

With increasing complexity of network services currently considered in network function virtualization~(NFV), it becomes increasingly crucial to clearly specify their structure and behavior based on their constituting virtual network functions~(VNFs). 
Such a clear specification is required to avoid ambiguity and provide all relevant information for
orchestration with Quality of Service~(QoS) guarantees.

\begin{figure}[tb]
	\centering
	\includegraphics[width=1\columnwidth]{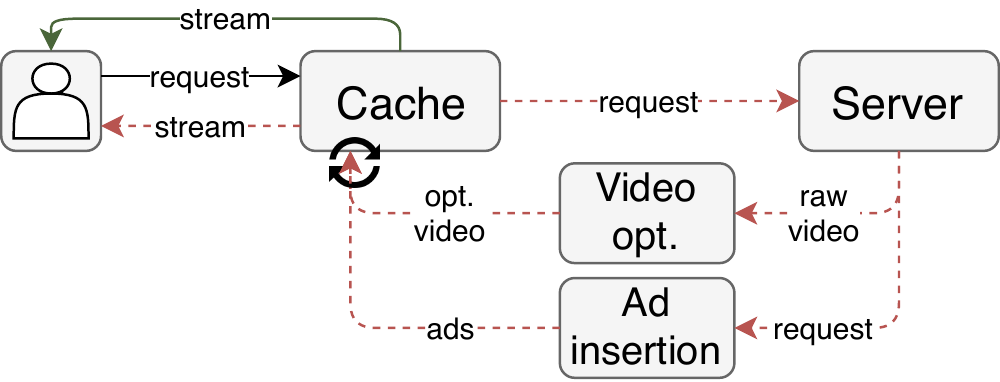}
	\caption{Example video streaming network service with stochastic behavior (cache hit/miss), synchronization (of video and advertisements), and loops.}
	\label{fig:example}
\end{figure}

To illustrate this requirement, we use the example in Fig.~\ref{fig:example}, showing a video streaming network service. Such video streaming network services are a common use case of virtual content delivery networks~(vCDNs)~\cite{Etsi13usecase}. In this network service, users request video streams from a vCDN cache. If the requested videos are available at the cache (cache hit), they can be streamed directly back to the users (solid, green arrow). In case of a cache miss, the request is forwarded to the storage server in order to load the requested videos to the cache (dashed, red arrow).
After retrieving the videos from the server, they are optimized by the
video optimizer (e.g.,~for mobile video). In parallel, the server sends
a request to an ad insertion VNF that selects suitable advertisements to be inserted in the video. The optimized video and the selected advertisements have to be synchronized at the cache before streaming the requested video (with advertisements) to the users.
An alternative example are social media overlays that are synchronized with streamed videos, showing live updates from a social network within the video (e.g., user comments during a sports match).

Current specification techniques are ambiguous as they do not clarify whether incoming traffic at multiple inputs of a VNF is synchronized or not. Synchronization can lead to additional delays, which need to be considered during orchestration to ensure QoS guarantees.
While there has been a lot of progress in NFV-related research and the example network service is quite simple, none of the existing specification techniques can properly specify and express its behavior and structure.
Existing approaches (e.g. ETSI~\cite{Etsi14mano}) cannot express synchronization needs of traffic from multiple connections, such as the synchronization of video and corresponding advertisements at the cache in Fig.~\ref{fig:example}.

Similarly, current specification techniques are ambiguous regarding traffic characteristics and the assignment of outgoing traffic to the outputs of a VNF, which are essential for orchestration.
For example, they cannot distinguish the following two cases: The cache in Fig.~\ref{fig:example} sends out \emph{either} a video stream to the users \emph{or} a request to the server, leading to either low or high end-to-end delay, respectively. In contrast, the server sends out VNF requests to \emph{both} the video optimizer \emph{and} the ad insertion.
Considering such individual VNF requests rather than just the overall traffic volume is necessary to estimate the VNF workload and required resources as well as the end-to-end delay. This information is crucial for successful orchestration, especially with QoS guarantees.

Finally, existing approaches assume network services to have a relatively simple, unidirectional structure. As stated by IETF~\cite{Ietf15, Ietf15Architecture}, common network services often have loops and are bidirectional, i.e.,~some VNFs are traversed multiple times. In the example of Fig.~\ref{fig:example}, requested videos return to the users (bidirectional) and the cache is traversed again when loading videos from the server.
Network services with such structures cannot be specified at all with current approaches and thus cannot be efficiently deployed and managed by management and orchestration~(MANO) systems.

Our main \textbf{contribution} is to fill the identified gaps by proposing an approach using Petri nets for clearly and formally specifying network services and the behavior of involved VNFs. 
Petri nets are an established method to express and analyze the behavior of distributed and concurrent processes (Sec.~\ref{sec:background}) and have been used successfully in related areas (Sec.~\ref{sec:relatedwork}).
Furthermore, they allow a clear and unambiguous specification of the behavior at the inputs and outputs of a VNF (e.g.,~synchronization) and allow the specification of any network service structure (e.g.,~bidirectional with loops), solving the aforementioned problems.
Furthermore, we use Petri nets to formally specify queuing, stochastically distributed processing delays, and changing traffic volume and characteristics at each VNF, accurately capturing the behavior of VNFs and the overall network service (Sec.~\ref{sec:specification}). The approach is not limited to VNFs but also works with physical network functions.
Specifying network services with Petri nets also enables diverse analysis options, allowing developers to perform extensive performance evaluations (Sec.~\ref{sec:analysis}).
In Sec.~\ref{sec:mano}, we discuss the implications of the specified behavior for placement and orchestration of network services and outline how MANO systems can leverage the analysis possibilities of Petri nets to ensure QoS guarantees.



\section{Background on Petri nets}\label{sec:background}
This section summarizes core ideas of Petri nets, as a basis for the
following discussions. 

A Petri net~\cite{Murata89} is a specific kind of directed, weighted, and bipartite graph comprising two kinds of nodes: 
\emph{Transitions}, modeling active components, and \emph{places}, describing passive entities. 
Directed \emph{arcs} connect either places to transitions or transitions to places and typically model the flow of data in the system, i.e., a transition consumes data (from places) and produces data (on places). 
In graphical representations, places are depicted by circles and transitions by boxes or bars. 

More formally, a Petri net can be defined as a 5-tuple $PN = (P, T, A, W, M_{0})$ with 
\begin{itemize} 
	\item $P$ the finite set of places,
	\item $T$ the finite set of transitions,
	\item $A \subseteq (P \times T) \cup (T \times P)$ the set of directed arcs,  
	\item $W: A \rightarrow \mathbb{N}$ the weight function, and
	\item $M_{0}: P \rightarrow \mathbb{N}$  the initial marking. 
\end{itemize}
The sets of places and transitions have to be disjoint.
The input places~$P^\mathrm{in}_t$ of a transition~$t$ are all places with arcs going into $t$. Similarly, the output places~$P^\mathrm{out}_t$ are all places to which $t$ has an outgoing arc.

In Petri nets, data is abstractly represented by \emph{tokens}, which reside on places. 
A marking describes how many tokens each place currently contains, where the initial marking~$M_0$ is the first marking.
Depending on the current marking and the weight function~$W$, an activity can occur, i.e., a transition consumes tokens from its input places and adds tokens at its output places.
Specifically, a transition~$t$ is enabled if each of the input places~$p \in P^\mathrm{in}_t$ has at least $W(p,t)$ tokens. If $t$ is enabled, it can fire, removing $W(p,t)$ tokens from every input place $p \in P^\mathrm{in}_t$ and adding $W(t,p')$ tokens to every output place $p'\in P^\mathrm{out}_t$.
Note that transitions without input places are always enabled and generate new tokens periodically.

\begin{figure}[tb]
	\centering
	\includegraphics[width=0.6\columnwidth]{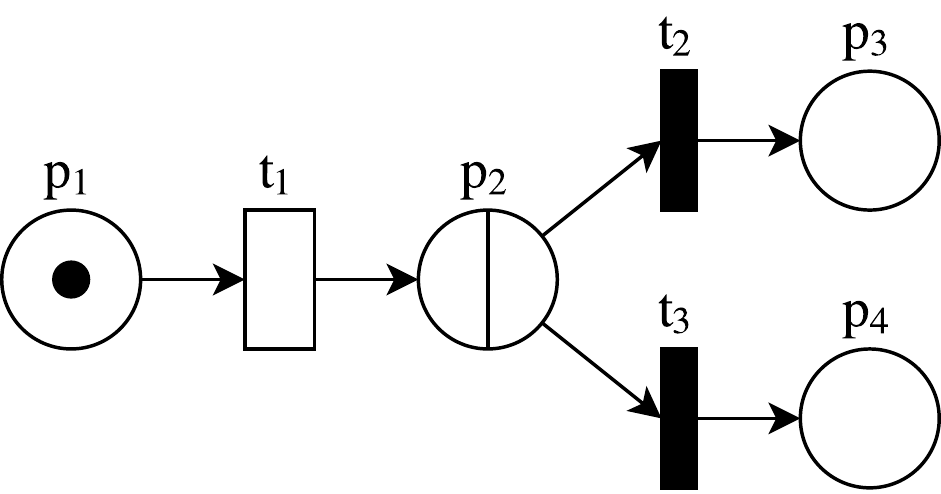}
	\caption{Example of a simple Queuing Petri Net.}
	\label{fig:qpn-example}
\end{figure}

For ease of modeling, this basic concept of Petri nets has been extended to various high-level Petri nets. We propose to use Queuing Petri Nets~(QPNs)~\cite{Bause93} for specifying and analyzing network services.
QPNs combine and extend stochastic~\cite{Ajmone84} and colored~\cite{Jensen87} Petri nets, supporting transition timing, token colors, and queuing places. We describe each of these features in the following, using the example in Fig.~\ref{fig:qpn-example}.

\emph{Transition timing:} The set of transitions~$T$ is divided into immediate and timed transitions. In Fig.~\ref{fig:qpn-example}, $t_1$ is timed and $t_2$ and $t_3$ are immediate.
While immediate transitions fire directly when enabled, timed transitions only fire after a certain delay~$d(t)$. This delay can either be fixed or follow a stochastic distribution (e.g., a uniform distribution with $d(t) = \mathrm{Uni}(1, 2)$).
If two or more immediate transitions are enabled, one of the transitions is selected randomly to fire first. The firing weight~$f(t)$ determines the probability for each transition to be selected. 

\begin{figure}[tb]
	\centering
	\includegraphics[width=0.3\columnwidth]{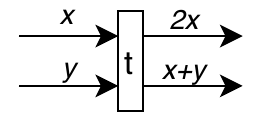}
	\caption{Transition~$t$ changes the color (here: numerical values) of traversing tokens as indicated by the arc expressions.}
	\label{fig:arc-expressions}
\end{figure}

\emph{Token color:} To distinguish different tokens and to convey additional information, tokens can take arbitrary values (or tuples of values) called colors. 
When tokens traverse a transition, the transition may manipulate the token color. In QPNs, arc expressions specify how a transition changes the color of traversing tokens. 
For each outgoing arc~$a_\mathrm{out}$, $E(a_\mathrm{out})$ defines how the token color is changed in relation to the color of incoming tokens.
In the example of Fig.~\ref{fig:arc-expressions}, token colors are numerical values, where the value of tokens from the arc referenced as $x$ is doubled at the first output. At the second output, outgoing tokens have the aggregated value $x+y$ of incoming tokens from both inputs.

\emph{Queuing places:} In a QPN, there are ordinary place and queuing places. While tokens in ordinary places simply form an unordered set, queuing places organize tokens in queues with a certain access strategy (typically FIFO). 
When firing, transitions take tokens from the front of the queues of their input places and generate tokens at the end of the queues at their output places.
In Fig.~\ref{fig:qpn-example}, $p_1, p_3, p_4$ are ordinary places and
$p_2$ is a queuing place.


\section{Specifying network services using QPNs}\label{sec:specification}

To formally and accurately specify network services and the behavior of their constituting VNFs, we use Queuing Petri Nets~(QPNs) as defined in Sec.~\ref{sec:background}.
In the following, we describe how developers can model traffic characteristics and systematically specify the behavior of the individual VNFs, taking queuing, synchronization of incoming traffic, processing, etc. into account. 
Based on the specified behavior of the individual VNFs, we illustrate how to specify an overall network service using the video streaming example of Fig.~\ref{fig:example}.
The resulting specification is useful for accurately analyzing the network service's behavior and performance (Sec.~\ref{sec:analysis}) and can support MANO systems in selecting service placements with better QoS (Sec.~\ref{sec:mano}).

\subsection{Traffic characteristics}\label{sec:sources}
We use tokens to model generic units of traffic, e.g., representing VNF requests or corresponding responses. These tokens can convey any kind of information in their token color, e.g., the request type and the size of the request in bytes. When a token is processed by a VNF, the VNF can behave differently for different token colors and can modify the color as further discussed in Sec.~\ref{sec:color-changes}.

\subsection{Network function behavior}\label{sec:spec-nf}
Using QPNs, it is possible to clearly and formally define the relevant behavior of VNFs that are involved in a network service. 
Behavior like queuing, synchronization, traffic processing, and different options for assigning traffic to the outputs greatly impact the overall behavior of the network service and should be considered during specification.
In the following, we describe each of these aspects and explain how to specify them with QPNs.

\subsubsection{Queuing}\label{sec:queuing}
When tokens arrive at a VNF and cannot be processed immediately, they are queued. In QPNs, this behavior can be specified using places at the inputs of a VNF. Using queuing places, a specific queuing strategy is applied to queue tokens before they become accessible to the VNF. We assume FIFO (first in, first out) queuing, but other queuing strategies can also be specified. The use of queuing places ensures that tokens traverse VNFs in an ordered way and captures potential queuing delays.

\subsubsection{Synchronization}\label{sec:synchronization}
If a VNF has multiple inputs, it is fundamental to distinguish whether
the traffic from different incoming arcs is synchronized or
not. In the example network service in Fig.~\ref{fig:example}, videos
and corresponding advertisements are synchronized at the cache to produce
complete video streams. The cache queues incoming videos until corresponding advertisements are available (or vice versa) and can be synchronized, possibly inducing additional delays.
If the traffic at different inputs is
independent of each other, no synchronization is needed and the tokens at different inputs
do not have to wait for each other. The distinction is important as it
affects the end-to-end delay and throughput of a network service and, hence, has to
be taken into account during orchestration or evaluation.

\begin{figure}[tb]
	\centering
	\subfloat[\label{fig:sync}Synchronization]{\includegraphics[width=0.45\columnwidth]{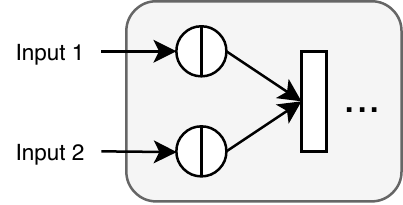}}
	\hfil
	\subfloat[\label{fig:nosync}No synchronization]{\includegraphics[width=0.47\columnwidth]{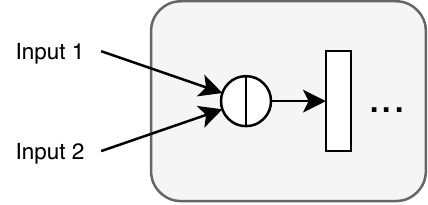}}
	\caption{QPNs can be used to easily specify whether the inputs of a VNF are (a)~synchronized or (b) not.}
	\label{fig:synchronization}
\end{figure}

To specify synchronization in a QPN, separate queuing places are used for each
input (Fig.~\ref{fig:sync}). Each of these input places has an arc to the timed transition
representing the VNF. This indicates that the transition can only fire once there is a token at each input place, i.e.,~the VNF synchronizes the traffic from all inputs before processing it.
Weight function~$W(p,t)$ specifies how many tokens at each input place~$p$ are required for synchronization, which are then consumed by transition~$t$ (typically, one per input).

If no synchronization is required, a single queuing place can be used
for all inputs, i.e.,~the incoming arcs are connected to the same input place (Fig.~\ref{fig:nosync}). This place is directly connected to the timed transition of the VNF. In doing so, incoming tokens are collected in a single queue and can be processed directly by the VNF independent of tokens from other inputs.
For VNFs with more than two inputs, it is also possible to only synchronize some but not all inputs by combining the different specification options.

\subsubsection{Processing delay}\label{sec:processing-delay}
In QPNs, the core element of a VNF~$i$ is a timed transition~$t_i$ that represents the processing of the VNF. The processing delay of the VNF can be specified by setting delay~$d(t_i)$.
This delay can represent the abstract effort for processing (e.g., number of instructions) or the actual processing time depending the underlying hardware (e.g., based on benchmarking results~\cite{Ietf17benchmarking, Ietf18benchmarking}). 
The processing delay of a VNF can either be fixed or follow a stochastic distribution. 
It can also be a function of the token color, e.g., to express different processing times for different request types.

\subsubsection{Token color changes}\label{sec:color-changes}
As mentioned before, the token color can represent multiple attributes of traversing traffic such as request type (e.g., HTTP request) and size in byte.
When processing tokens, a VNF can change any attribute of the token color such as the traffic volume. For example, WAN optimizers compress traffic and reduce traffic volume, but BCH encoders increase the volume~\cite{Ma17}. In QPNs, the color of outgoing tokens in relation to incoming tokens (possibly at multiple inputs) can be expressed using arc expressions~$E$.
%

\subsubsection{Token number changes}\label{sec:number-changes}
In addition to the token color, transitions also decide the number of outgoing tokens as specified by weight function~$W$. For example, multiple incoming tokens can be merged to a single outgoing token or an incoming token can be replicated into multiple outgoing tokens.

\subsubsection{Selection of outputs}\label{sec:output-selection}
If a VNF~$i$ has only one output, its transition~$t_i$ is directly connected to the next VNF such that its outgoing tokens directly enter an input place of the next VNF.
If a VNF has multiple outputs, it may want to send requests on all outputs. For example, the server in Fig.~\ref{fig:example} sends requests to both the video optimizer and the ad insertion.
For each output, the number of tokens and their individual color is specified using separate weights and arc expressions, respectively.

\begin{figure}[tb]
	\centering
	\includegraphics[width=0.6\columnwidth]{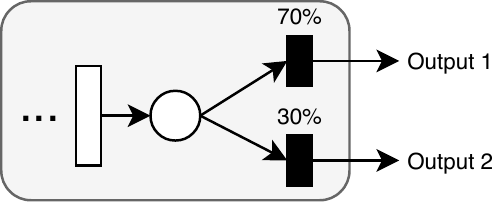}
	\caption{One of two outputs is selected randomly according to the firing weights of the immediate transitions (here, 70\% and 30\%).}
	\label{fig:output-selection}
\end{figure}

Alternatively, a VNF with multiple outputs may select a subset of its outputs whenever sending out tokens, e.g., to perform load balancing. Randomly selecting one output for each outgoing token can be specified as follows in QPNs: An additional place collects outgoing tokens and immediate transitions are connected for each output. The firing weight~$f(t)$ determines the probability for each of the immediate transitions to fire first, assigning the token to the corresponding output.
Fig.~\ref{fig:output-selection} shows an (incomplete) VNF, where one of two outputs is randomly selected.
It is also possible to express other selection methods such as round robin by adding auxiliary places and transitions (for brevity, we skip details).

Finally, the selection of outputs may depend on the token color, e.g., leading to varying selection probabilities of the outputs. 
To specify such general behavior, immediate transitions could be extended to support varying firing weights that depend on the token color.
Hence, in the most general case, a VNF has a list of production rules that depend on the incoming token color. Each production rule generates an outgoing token, specifying the token color and the probability distribution for the different outputs to be selected.

\subsection{Complete network services}\label{sec:complete-ns}
The presented comprehensive model allows a flexible, yet clear and unambiguous specification of the behavior of individual VNFs inside a network service.
Having all involved VNFs specified formally using QPNs, they can easily be chained to form a complete network service by connecting the outgoing arcs of a VNF with the input places of the next VNF.
In doing so, any kind of structure can be specified, including uni- or bidirectional network services and even network services with multiple loops.
When specifying network services with loops, where some VNFs are traversed multiple times, additional places and transitions can be added to separate the behavior for each traversal (e.g., for upstream and downstream).

To represent sources (e.g., users or sensors) and specify how they are connected to the network service, timed transitions can be added in front of the first VNF.
In QPNs, such timed transitions without input places periodically generate tokens of a certain color and at a certain rate. This rate can correspond to fixed or randomly distributed inter-arrival times of tokens (e.g., forming a Poisson process), depending on the timing~$d(t_s)$ of the corresponding transition~$t_s$. To specify the generation of tokens with different colors (e.g., different types of requests), multiple transitions can be used.

After traversing all subsequent VNFs of a specific network service, the tokens finally
reach an end-point, where the tokens are consumed. In QPNs, such
end-points are modeled using places, collecting the incoming
tokens. In bidirectional network services, where traffic returns to
the sources, the places representing end-points are co-located with
the transitions representing sources.

\subsection{Example: Video streaming service}\label{sec:qpn-video-example}
\begin{figure}[tb]
	\centering
	\includegraphics[width=1\columnwidth]{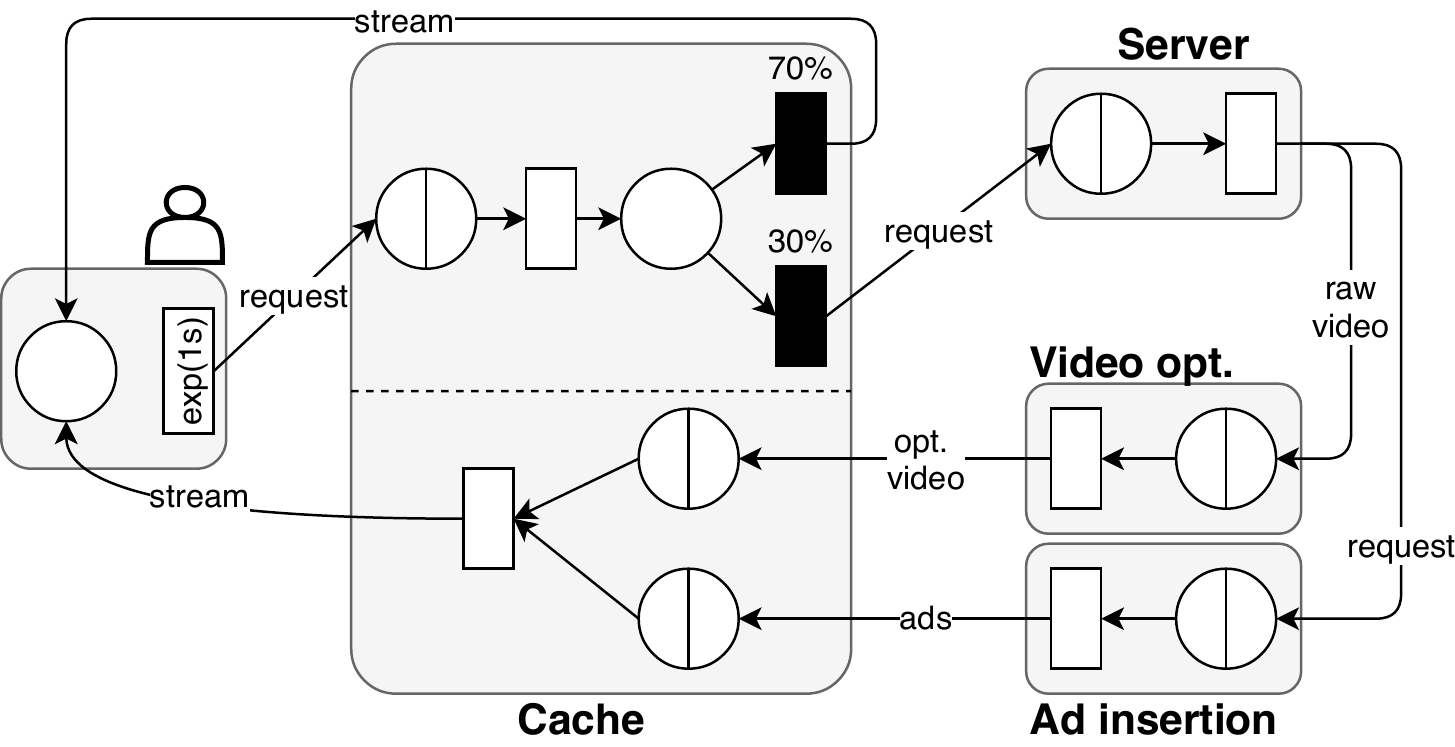}
	\caption{Video streaming network service of Fig.~\ref{fig:example} specified as QPN.}
	\label{fig:pn-example}
\end{figure}

We illustrate the aforementioned possibilities of QPNs by specifying the example video streaming network service of Fig.~\ref{fig:example}. In the resulting QPN in Fig.~\ref{fig:pn-example}, all places and transitions belonging to one VNF are visually framed with a gray, labeled rectangle (not part of the QPN notation).

To illustrate the specification of sources and end-points, a user component is modeled using a transition and a
place. The transition generates new tokens, representing requests for video streams of a certain size, e.g., in exponentially distributed time intervals with mean 1~second.
The place at the user is the end-point of the network service and collects returning tokens (i.e., the requested video streams).
Other specifications are also conceivable, e.g.,~where users only request new video streams after receiving (and watching) previously requested ones. This could easily be specified by connecting the place to the transition.

Incoming user requests at the cache are queued at the input place and then traverse the timed transition with delay corresponding to a cache lookup. Depending on the success of the lookup, requests reach one of the two immediate transitions with a probability corresponding to the cache hit ratio. Here, we assume a cache hit ratio of 70\% and specify it using firing weights. In case of a cache hit, the token traverses the upper transition and returns to the user. When traversing the transition, it changes the color of the token to represent a returning video stream with much larger size than the incoming request.

In case of a cache miss (here with probability 30\%), the token is forwarded via the lower immediate transition to the queuing place of the server. After processing the incoming request, the server transition generates two outgoing tokens of different color: One token representing the raw video, going to the video optimizer, and one token requesting ad insertion from the corresponding VNF.

The video optimizer also delays traversing tokens corresponding to its processing delay and modifies the token color according to the optimized bit rate of the videos. At the same time, tokens traverse the ad insertion, where they are processed and the outgoing tokens represent the selected advertisements. The tokens then reach the two downstream inputs of the cache, which are visually separated from the upstream direction by a dashed line (not part of the QPN notation). At the cache, the video and advertisements are synchronized. 
Only when both tokens are available at the transition, it generates a new outgoing token representing the complete video stream. While this synchronization can introduce additional delay, it is necessary to ensure that each video is matched with the corresponding advertisements.

The example illustrates the capabilities of specifying complex network services using QPNs. Each involved VNF and its behavior is described clearly (e.g.,~synchronization at the cache). The specification remains simple, using only few places and transitions, especially for VNFs with just one input and output such as the server.
While the bidirectional structure of the video streaming network service (containing backward loops) cannot be expressed with current specification approaches, it can easily be specified with QPNs.

\section{Analyzing QPN-specified network services}\label{sec:analysis}
As QPNs are an established method, which has been used for years, different tools are available supporting the specification of QPNs.
Among these tools, TimeNet~4.4~\cite{timenet} stands out since it supports most features for formally specifying
network services introduced in Sec.~\ref{sec:specification}, e.g., synchronization, stochastically timed transitions, arbitrary token colors, and firing weights. Developers can use its graphical user interface~(GUI) to specify their network services in a simple drag-and-drop manner.

In theory, network services specified as QPNs can be analyzed mathematically, e.g., to compute the steady state distribution or check for deadlocks. In practice, such theoretical analysis is often intractable due to explosion of possible states, leading to prohibitive complexity~\cite{Bause96Stochastic}.
Hence, we focus on efficient analysis of such network services through simulation.

Since TimeNet~4.4 and similar tools only have limited analysis support, we provide a simulation compiler that generates simulation code based on the QPN specification for the popular simulation framework OMNeT++~\cite{omnet} (Sec.~\ref{sec:compiler}). 
In Sec.~\ref{sec:case-study}, we illustrate some of the various analysis possibilities in a case study.
We provide the compiler software and the analysis case study as publicly available open-source project~\cite{github}. In doing so, we hope to encourage further experimentation and contributions.

\subsection{Generating discrete event simulations}\label{sec:compiler}
To alleviate TimeNet's limited simulation capabilities, we introduce a compiler that processes
TimeNet's structured XML-based storage format to automatically create simulation programs
for general-purpose simulation tools. As an example, the compiler generates simulations
for OMNeT++~5.2 that realize the same semantics as the QPNs specified with TimeNet.

Specifically, the compiler automatically generates modules in OMNeT++ corresponding to the different places and transitions. 
It then connects the modules to form a network service as specified in
TimeNet. The compiler considers all specified characteristics like
processing delay, queuing, synchronization, and changes of token color
by retrieving them from the XML specification and automatically
mapping them to configuration files in OMNeT++.
The generated files can be used directly as an OMNeT++ project to accurately simulate and analyze the behavior of the specified network service.

During the simulation, OMNeT++ measures
and collects various metrics such as end-to-end delay of tokens as well as delay to each VNF, token
color, processing delays, and queue lengths. 
The compiler can easily be extended, e.g., to include other measurements or to integrate generation backends for other simulation frameworks or even emulation platforms.

\subsection{Case study: Analyzing simulation results}\label{sec:case-study}
After specifying a network service as QPN with TimeNet and generating the corresponding simulation code, developers can accurately and efficiently simulate the behavior of their network service. 
OMNeT++ can simulate the behavior of the network service over arbitrary time intervals and with different random seeds, providing results with high level of confidence. It also supports automated parameter studies, where different configurations are tested automatically while performing fine-grained measurements. 
The proposed workflow is highly automated and time-efficient. For example, the simulation code for the video streaming network service of Sec.~\ref{sec:qpn-video-example} generates in milliseconds and its behavior over ten simulated hours can be recorded in less than a second. 
Hence, developers can quickly and easily test different possible configurations or implementation options of their network service and the involved VNFs. 

\begin{figure}[tb]
	\centering
	\includegraphics[width=0.7\columnwidth]{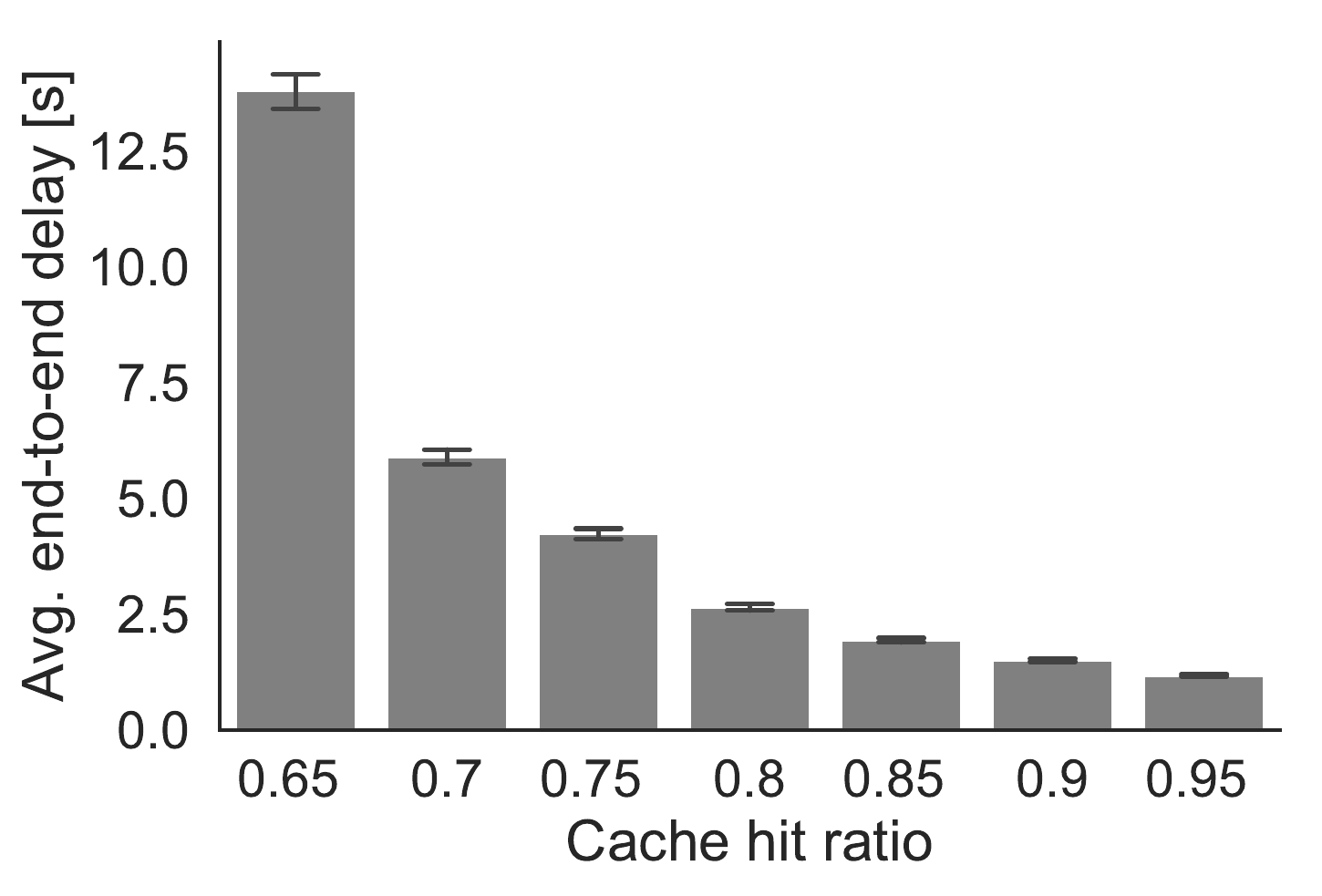}
	\caption{Impact of different cache hit ratios on the average end-to-end delay.}
	\label{fig:cache-hit}
\end{figure}

For example, in the video streaming network service, the cache hit ratio has a crucial influence on the service's end-to-end delay. For every cache miss, the requested video needs to be loaded from the server and processed by the video optimizer and ad insertion, leading to much higher end-to-end delay than a cache hit. 
Simulating and analyzing different cache configurations with different cache hit ratios enables a better understanding of their impact on the end-to-end delay.
Fig.~\ref{fig:cache-hit} shows the average end-to-end delay in relation to different cache hit ratios based on 30 simulation runs with random seeds. Here, the analysis shows that the average end-to-end delay quickly decreases with an increasing cache hit ratio. This is not surprising as a higher ratio of cache hits means that more videos can be streamed directly from the cache with short end-to-end delay. However, the figure also indicates that further improving an already high cache hit ratio only minimally reduces the end-to-end delay. Here, changing from a 65\% to a 70\% cache hit ratio yields the biggest improvement in average end-to-end delay. 
Such insights help developers to recognize which components benefit most from further improvements.

\begin{figure}[tb]
	\centering
	\includegraphics[width=0.7\columnwidth]{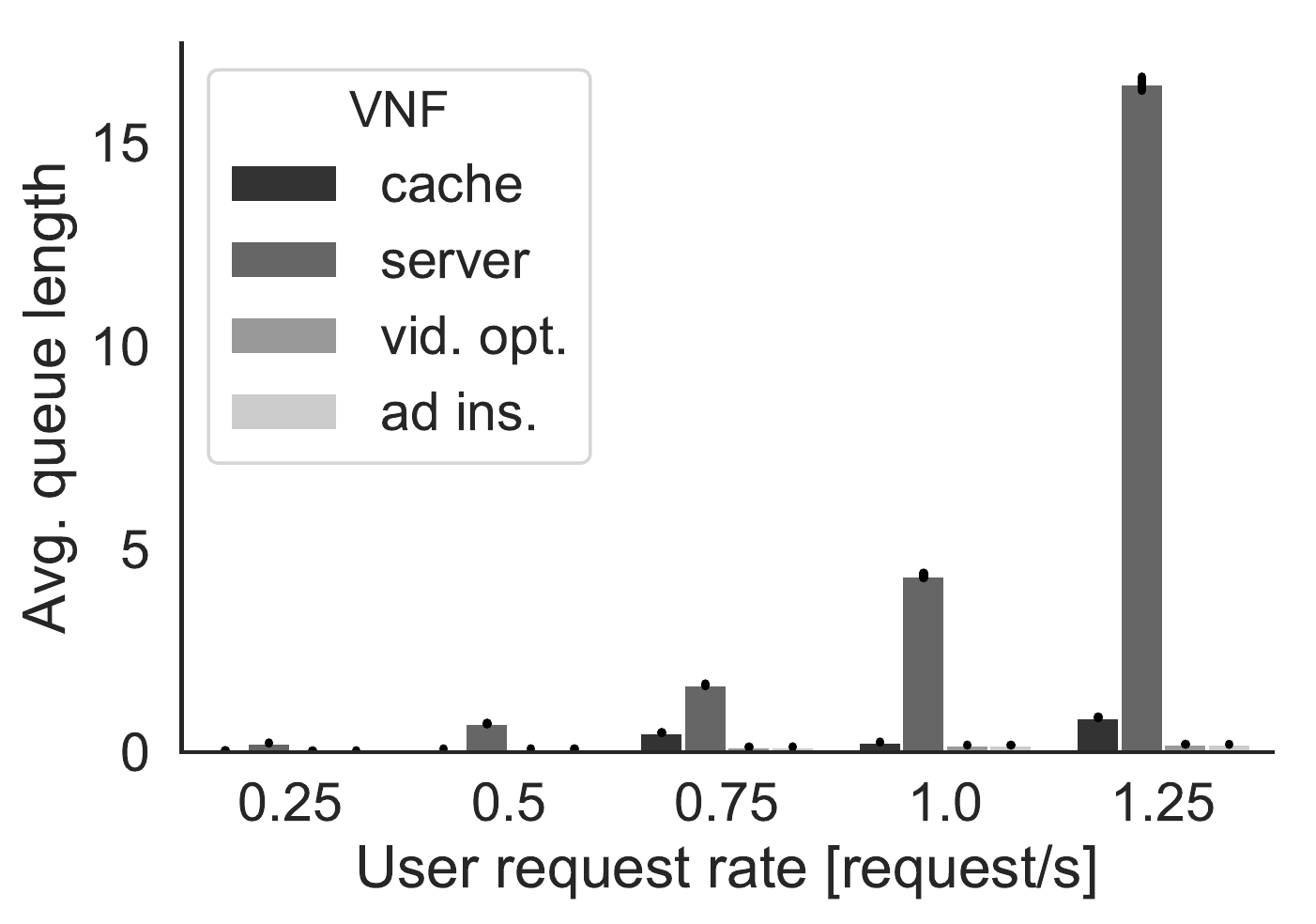}
	\caption{Queue lengths of each VNF with increasing user request rate.}
	\label{fig:bottleneck}
\end{figure}

In addition to analyzing the impact of different configurations, it is also useful to investigate possible bottlenecks of a network service in order to ensure QoS even for high load. 
To identify a bottleneck, the network service is simulated multiple times with an increasing rate of user requests. As long as the rate is not too high, incoming requests are processed immediately and the VNFs' queues stay fairly empty. Once requests arrive at a higher rate than they can be processed, the queues start filling up and the end-to-end delay quickly increases over time.
Fig.~\ref{fig:bottleneck} shows the queue lengths of all involved VNFs for an increasing user request rate. The figure shows that, in contrast to the other VNFs, the server's queue length dramatically increases for higher request rates, which leads to high end-to-end delays. This indicates that the bottleneck is at the server since it cannot process higher request rates quickly enough.
Hence, developers may focus on a more efficient implementation of the server to decreasing its processing time and enable higher user request rates.

This case study illustrates some of the possible analysis options when specifying network services as QPNs and generating corresponding simulation code with the provided compiler. Further metrics can easily be measured and analyzed to enable a more in-depth analysis. Overall, the described specification and analysis workflow supports developers in understanding and improving their network services.


\section{Placing QPN-specified network services}\label{sec:mano}
\begin{figure}[tb]
	\centering
	\includegraphics[width=1\columnwidth]{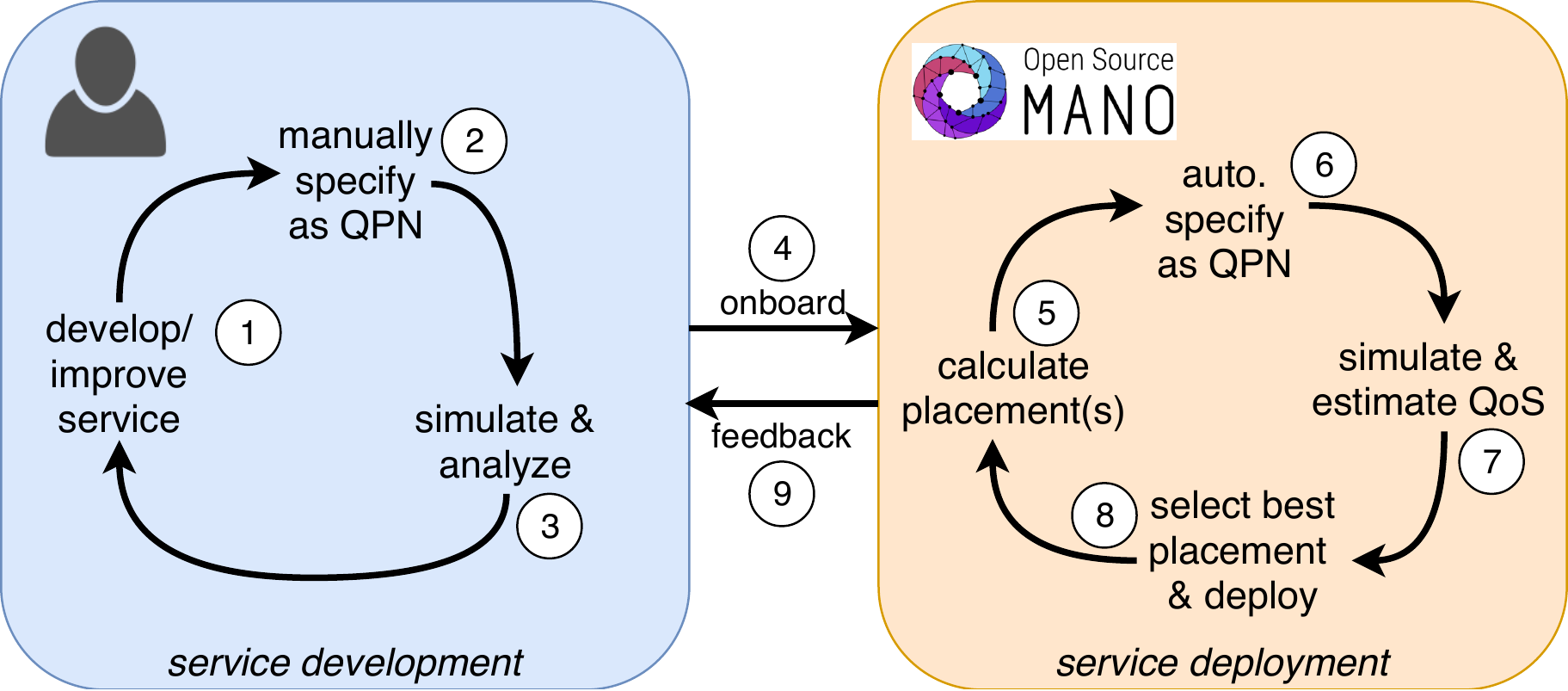}
	\caption{Envisioned DevOps approach: Both developers and MANO systems benefit from using QPNs.}
	\label{fig:devops}
\end{figure}

The expressive specification of network services as QPNs and their extensive analysis capabilities also provide benefits when deploying network services through a MANO system. 
We envision a DevOps approach, in which both developers and MANO systems profit from using QPNs (Fig.~\ref{fig:devops}). 
After developing their network service, developers specify its behavior using QPNs and analyze its performance through simulation as described in Sec.~\ref{sec:specification} and Sec.~\ref{sec:analysis}, respectively (steps~1-3 in Fig.~\ref{fig:devops}).
Once satisfied with the service, developers onboard it to a MANO system such as OSM~\cite{osm} for deployment (step~4).

When deploying a network service, the MANO system needs to place the network service in the underlying substrate network. This requires efficient mapping of the involved VNFs to network nodes and interconnecting them along network links.
In Sec.~\ref{sec:mano-spec}, we outline how MANO systems can leverage the expressive specification of QPN-based network services to calculate better, more informed placements (step~5).
In Sec.~\ref{sec:mano-analysis}, we argue that even without using the QPN specification for calculating improved placements, MANO systems can still benefit from the accurate analysis possibilities of QPNs to decide between different placement options (steps~6-9).
While we do not propose specific orchestration algorithms for QPN-specified network services, we highlight the impact and benefits of having a QPN specification available.

\subsection{Leveraging QPN specifications for placement}\label{sec:mano-spec}
Even though network services are often bidirectional or contain loops~\cite{Ietf15, Ietf15Architecture}, typical service descriptors cannot specify such network services. In contrast, QPNs allow to specify network services with arbitrary structures (e.g., containing loops). 
Hence, extending MANO systems to understand QPN-specified network services opens the door to support placement of realistic, bidirectional services with loops.

Even when placing network services without loops, using the additional information of a QPN specification allows a more informed placement decision. For example, using QPNs, developers can clearly specify whether incoming traffic from multiple VNFs is synchronized or not, whereas typical descriptors are ambiguous. 
If the traffic from multiple VNFs is synchronized, it can lead to additional synchronization delay, which should be minimized during placement. 

For example, in the video streaming service of Fig.~\ref{fig:example}, videos and advertisements coming from the video optimizer and ad insertion, respectively, are synchronized at the cache. 
Hence, an orchestration algorithm should try to place these VNFs such that both the video optimizer and the ad insertion are equally close to the cache. In doing so, videos and corresponding advertisements arrive at roughly the same time at the cache, minimizing the synchronization delay.
In contrast, placing only the video optimizer (or only the ad insertion) close to the cache would not help. For synchronizing videos with advertisements, videos arriving at the cache would be blocked until the corresponding advertisements arrive, leading to additional delay.

Generally, MANO systems can leverage the QPN specification of a network service to better understand its behavior and anticipate the resulting QoS of possible placements. Placement algorithms taking this information into account can compute better, more informed placements, which is especially beneficial for QoS-sensitive network services.

In practice, the QPN specification of a network service (e.g., the XML format produced by TimeNet~\cite{timenet}) may be referenced in an additional, optional field within the typical service descriptors (e.g., ETSI's network service descriptor~\cite{Etsi14mano}). In doing so, MANO systems supporting QPN specifications could leverage the detailed specification while others could simply ignore it and use only the standard information included in the descriptors.

\subsection{Leveraging QPN analysis for QoS estimates}\label{sec:mano-analysis}
Even without using specific placement algorithms that leverage the detailed information of QPN specifications, MANO systems can still benefit from the powerful analysis options of QPNs by estimating and comparing the QoS of possible placements (e.g., regarding end-to-end delay).

Using a regular placement algorithm, a MANO system can compute a set of promising placements based only on the typical service descriptors (step~5 of Fig.~\ref{fig:devops}). 
For each of the possible placements, the MANO system creates a matching QPN specification that represents the behavior of the placed network service (step~6). To this end, it automatically copies and adjusts the service's QPN specification provided by the developer. 
For example, it can add additional sources or adjust their request rate by modifying the delay of the corresponding timed transitions. Link delays between VNFs placed at different network nodes can be specified by adding a timed transition with the corresponding delay. Developing a specific algorithm for automating the QPN specification of placed services remains future work.

The QPN-specified placements can then be simulated automatically as described in Sec.~\ref{sec:analysis} (step~7). This allows MANO systems to check and evaluate the expected QoS based on the specified behavior, taking stochastic processing delays, queuing, synchronization, etc. into account.
In doing so, the MANO system can quickly simulate multiple promising placements to compare their estimated QoS and select the best one for deployment (step~8).

Hence, even without using the QPN specification when computing a placement, simulating the behavior of the placed service helps avoid unexpected high delays and violations of QoS guarantees.
To complete the DevOps cycle, collected measurements from the actual service deployment could be provided as feedback to the developer (step~9). Comparing the actual service behavior and QoS with the simulated one, allows to iteratively improve and fine-tune the QPN specification and the service itself.


\section{Related work}\label{sec:relatedwork}

\subsection{Related work on Petri nets}\label{sec:relatedwork-petri}
Similar to our approach of using QPNs to specify and analyze network services, QPNs have been used for specification and analysis in related scenarios. 
Kounev and Buchmann~\cite{kounev2003performance} discuss the benefits of QPNs compared to conventional modeling paradigms such as queuing networks. In a case study, they use QPNs to model a distributed e-business system with concurrent system behavior, synchronization, and blocking, which can also occur in network services. The specific model of a e-business system cannot adequately model network services in NFV as it misses the concept of interconnected components like VNFs.
Nevertheless, their research suggests the applicability of QPNs to both hardware and software aspects of a system such as physical and virtual network functions. Also their accurate performance prediction encourages the use of QPNs for specification and analysis.

In a follow-up paper~\cite{kounev2006performance}, Kounev considers generic component-based systems, describing the general high-level workflow to specify such systems with QPNs. We mostly followed the proposed workflow to specify network services with QPNs. However, the paper focus more on technical, internal aspects of the components (e.g., number of cores) rather than considering how these components are connected. In contrast, we also focus on distinguishing and clearly specifying how VNFs are interconnected. We also consider more complex token colors with multiple dimensions such as request type and size.

In other related work~\cite{rygielski2014data, Rygielski2016}, the authors specify and analyze the performance of data center networks using QPNs (taking different SDN switching modes into account~\cite{Rygielski2016}). The authors specify different kinds of network nodes as QPN subnets, which is similar to our approach of specifying the behavior of VNFs with multiple places and transitions. As the authors focus on automatically generating QPNs from a domain-specific language, they only consider three different kinds of nodes (start, intermediate, and end nodes), where all nodes of a type have the same structure. In contrast, the structure of individual VNFs in a network service can vary greatly and cannot be predefined. Depending on how a VNF deals with incoming traffic, processes it, and assigns it to its outputs, various different specifications may be required.
Furthermore, the authors only consider simplistic token colors, which mostly remain unchanged during execution. In our approach, tokens represent various traffic attributes, which are changed frequently and in a fine-grained way as tokens traverse chained VNFs.

Overall, the discussed related work has many similarities to ours, but none of the existing approaches takes the specific characteristics of network services into account, e.g., interconnected VNFs and changing traffic characteristics.
Nevertheless, the successful specification and analysis in previous work illustrate the possibilities of QPNs, confirming their modeling power and expressiveness.


\subsection{Related work on NFV}\label{sec:relatedwork-nfv}

Herrera and Botero provide a recent survey of different resource allocation approaches in NFV~\cite{Herrera16}. For optimal resource allocation and placement of network services, a clear specification of the network services and involved VNFs is required, providing all relevant information. 
Models for resource allocation typically assume network services to be unidirectional and do not clearly specify their behavior, e.g., regarding synchronization. In contrast, our specification technique using QPNs allows a clear, precise, and formal specification of the behavior of network services and involved VNFs, even supporting bidirectional network services with loops.
 
Similar to our approach, several authors~\cite{Ma17, Ma17journal, Draxler16} take changing volume of traffic into account when traversing different VNFs. Additionally, our specification can take arbitrary further traffic characteristics into account, e.g., the request type. 

Moens and De Turck~\cite{Moens14} model hybrid scenarios with VNFs
and physical network functions. Our specification approach is also
applicable to both virtual and physical network functions.
Luzelli et al.~\cite{Luizelli15} consider fixed processing delays for each VNF. We also consider processing delays but allow both fixed delays as well as stochastically distributed processing delays.
We assume the distribution of processing delay for each VNF to be known. Otherwise, Lei et al.~\cite{Lei17} show that the distribution can be obtained using machine learning.

Outside the standardization approaches by ETSI~\cite{Etsi14mano} and
IETF~\cite{Ietf15}, there has been very little work focusing on proper specification of network services.
Mehraghdam and Karl~\cite{Mehraghdam16} propose a YANG data model
supporting a flexible order of VNFs within a network service, i.e.,
the order of some VNFs can be switched without affecting  overall
functionality. We assume this order of VNFs to be already decided when
specifying a network service using QPNs in order to enable a meaningful
analysis of the specified network service. By specifying multiple
versions of a network service with different order of VNFs, each
version could  be analyzed and compared, e.g., regarding end-to-end delay.

Other related work focuses on the specification of scalable service templates~\cite{Draxler16, Draxler18bjointsp}. Such service templates describe the network service structure and involved VNFs in a network service in a general way without specifying a fixed number of instances. Instead, the number of instances per VNF is scaled dynamically according to the current load. This dynamic scaling can help ensure good service quality and avoid waste of resources.
Network services specified with our specification technique can also be interpreted as scalable templates, where the behavior of each VNF instance is formally specified using QPNs. When scaling out new instances, we can clearly specify how traffic is balanced between multiple instances of the same VNF or whether incoming traffic from multiple instances is synchronized.


\section{Conclusion}\label{sec:conclusion}
In this paper, we presented a novel approach for specifying and analyzing network services in the context of NFV. By using QPNs, the behavior of a network service and constituting VNFs can be specified in a simple, yet precise fashion.
In the specification, we take queuing, synchronization, probabilistic behavior, and complex structures with loops into account.
The clear and formal specification with QPNs ensures unambiguity and provides MANO systems with relevant information for optimizing the placement of VNFs.
By taking stochastically distributed processing delays and potential queuing and synchronization delays into account, our approach allows an accurate analysis of a network service's end-to-end delay and other metrics of interest such as throughput or queue lengths. Our proposed tool-based and partly automated workflow simplifies and accelerates the specification, simulation, and analysis process.
It allows developers to test different service configurations and to detect possible bottlenecks in order to further improve their services.
We also outline how MANO systems can use our approach to estimate and compare the QoS of possible placements. In doing so, they can select the best placement and provide QoS guarantees.

In future work, we plan to further investigate placement and orchestration of QPN-specified network services. Specifically, we intend to design placement algorithms that leverage the formal QPN-based specification to provide better service quality.
Moreover, we plan to extend a MANO system to automatically specify calculated service placements as QPNs, simulate them to estimate the QoS, and deploy the best one. This would enable the envisioned fully automated DevOps workflow of specification, placement, and analysis.


\section*{Acknowledgments}

This paper has received funding from the European Union’s
Horizon 2020 research and innovation programme under grant
agreement No. 761493 (5GTANGO) and 671517 (SONATA), and from the German
Research Foundation (DFG) within the Collaborative Research
Centre “On-The-Fly Computing” (SFB 901).

\bibliographystyle{IEEEtran}
\bibliography{ref}

\newpage

\end{document}